\documentstyle[11pt,subfigure,psfig]{article}
\def\diameter{{\ifmmode\mathchoice
{\ooalign{\hfil\hbox{$\displaystyle/$}\hfil\crcr
{\hbox{$\displaystyle\mathchar"20D$}}}}
{\ooalign{\hfil\hbox{$\textstyle/$}\hfil\crcr
{\hbox{$\textstyle\mathchar"20D$}}}}
{\ooalign{\hfil\hbox{$\scriptstyle/$}\hfil\crcr
{\hbox{$\scriptstyle\mathchar"20D$}}}}
{\ooalign{\hfil\hbox{$\scriptscriptstyle/$}\hfil\crcr
{\hbox{$\scriptscriptstyle\mathchar"20D$}}}}
\else{\ooalign{\hfil/\hfil\crcr\mathhexbox20D}}%
\fi}}

\def\arcsec{\hbox{$^{\prime\prime}$}}
\def\utw{\smash{\rlap{\lower5pt\hbox{$\sim$}}}}
\def\udtw{\smash{\rlap{\lower6pt\hbox{$\approx$}}}}

\def\fm{\hbox{$.\!\!^{\rm m}$}}

\def\farcm{\hbox{$.\mkern-4mu^\prime$}}

\newcommand{\McLellan}{${\rm M^cLellan}$}

\title{\LARGE {\bf Vilnius CCD Photometry of Southern Clusters:\\
Some Preliminary Results} }
\author
{ Timothy Banks, D.J. Sullivan, M.C. Forbes, \\
\vspace{2mm}
\small{\it Department of Physics, Victoria University of Wellington, P.O. Box
600, Wellington}\\
\vspace{2mm}
and \\
\vspace{2mm}
R.J. Dodd, \\
\small{\it Carter Observatory, P.O. Box 2909, Wellington.}
}
\date{\footnotesize{First Received: May 1993. Resubmitted: October 1994.
Accepted: October 1994}}
\begin{document}
\maketitle

\begin{abstract}
\noindent The Vilnius Photometric system has been used with a CCD
system only once before this study.  Preliminary crowded field
reductions of Vilnius CCD observations of the star clusters NGC~2004
and NGC~4755 (the $\kappa $ Crucis cluster) are presented, demonstrating
the feasibility of using the filter set in this manner given the
resources available to Victoria University of Wellington.  The
standard (photo-electric) filters were used for the first time in
conjunction with a CCD system.  It is also the first time this filter
system has been applied to an extra-galactic star cluster (NGC~2004).
\end{abstract}

\section{Introduction}

The seven filter Vilnius system (see Strai\u{z}ys 1992a, Forbes 1993)
makes possible the purely photometric determination of the spectral
classes, absolute magnitudes, and metallicities of stars while also
correcting for interstellar reddening. This is facilitated by the
careful thought given to the positioning and widths of the filters
relative to the spectral features of all luminosity classes. For
example, the U filter measures the ultraviolet intensity below the
Balmer jump, while the P filter is placed on the jump itself, allowing
luminosity determinations for early type stars. The Z, V and S filters
coincide with features in late-type stars (see Figure 1 of Dodd,
Forbes~\& Sullivan 1993).  Peculiar stars, such as metal-deficient
Giants and blue Horizontal Branch stars, can be recognized using the
two and three dimensional classification schemes of the system
(Strai\u{z}ys 1992b), making the filter set well suited for the study
of star clusters.  Further details of the rationale in the design of
the filter set may be found in Strai\u{z}ys~\& Sviderskiene (1972).
Figure~\ref{figure:two} plots the response functions for the Vilnius
filters in combination with the Mount John University Observatory
(MJUO) CCD.

\begin{figure}[t]
\vspace{3cm}
\begin{center}
{\it Figure unavailable in machine readable form}
\end{center}
\vspace{3cm}
\caption{ {\bf Vilnius Filter Response for the TH7882
Chip:} Measured transmittances of the Vilnius filters, and the
Manufacturer's Quantum Efficiency specifications for the TH7882 CCD
used in this study (given as a dotted line), were combined to show the
filter responses with this chip. For a flat spectrum, the percentage
transmittances of the U, P, X, Y, Z and S filters relative to V are
61, 53, 74, 95, 106 and 87 respectively.  \label{figure:two}}
\end{figure}

In light of these capabilities, the idea of extending the system to
the southern hemisphere was first considered in 1985 at the Royal
Observatory Edinburgh, given that none of the already established
standard regions (e.g.\ Zdanavi\u{c}us~{\em et al.} 1969,
\u{C}ernies~{\em et al.} 1989, and \u{C}ernies~\& Jasevi\u{c}us 1992)
extended south of the celestial equator. This programme commenced in
1988 using the 61cm telescopes at MJUO, with the initial goal of
establishing standards near the South Celestial Pole and also bright
(V $ < $ 7) stars generally distributed south of -20 degrees. To date
some 90 cluster, 109 primary standard, and 225 secondary standard
stars have been observed at least once.  Further details on the
programme may be found in Forbes (1993) and Forbes, Dodd~\& Sullivan
(1993).  Given this network, and the availability of both an
established image reduction system and medium size telescope equipped
with a CCD, test images were acquired of the open (Galactic) cluster
NGC~4755 and the LMC young ``globular'' NGC~2004.

\section{Observations}

Vilnius photometry has been realised with a CCD only once before, in
the cursory trial of Boyle~{\em et al.} (1990) who used a
non-standard filter set. The objective of the current study was to
determine if useful observations could be obtained using the Vilnius
filters in conjunction with the MJUO CCD camera and associated
hardware.

UPYV observations were made on March 5 1993 with the 1m \McLellan\
telescope at MJUO ($\rm170^o$ 27\fm 9 East, $\rm43^o$ 59\farcm 2
South) in $ \sim 3 \arcsec $ seeing, using a cryogenically cooled
Thomson TH7882 CDA charge-coupled device.  This chip has 384 by 576
pixels. Each pixel is 23~$\mu$\/m across, which at the f/7.9
Cassegrain focus used by this study corresponds to 0.60\arcsec\ (Tobin
1989). Images were collected using the Photometrics PM-3000 computer
running FORTH (Moore 1974) software with extensive local
modifications, and written to half inch 9 track magnetic tape for
transportation back to Victoria University of Wellington (VUW) for
analysis. Images from these tapes were then converted into the FITS
(Wells, Griesen~\& Harten 1981) format from the native Photometrics
one, and read into the Image Reduction and Analysis Facility
(IRAF)\footnote{ Courtesy of the National Optical Astronomical
Observatories, which are operated by the Association of Universities
for Research in Astronomy under cooperative agreement with the
National Science Foundation.}, where subsequent reduction took place.
Details on the data pathway and Image processing facility established
at VUW are in Banks (1993).  Further details on the MJUO CCD data
acquisition system and its characteristics may be found in Tobin
(1992).

\subsection{NGC 2004}

\begin{figure}[t]
\begin{tabular}[t]{c}
\subfigure[]
 	{
	\psfig{figure=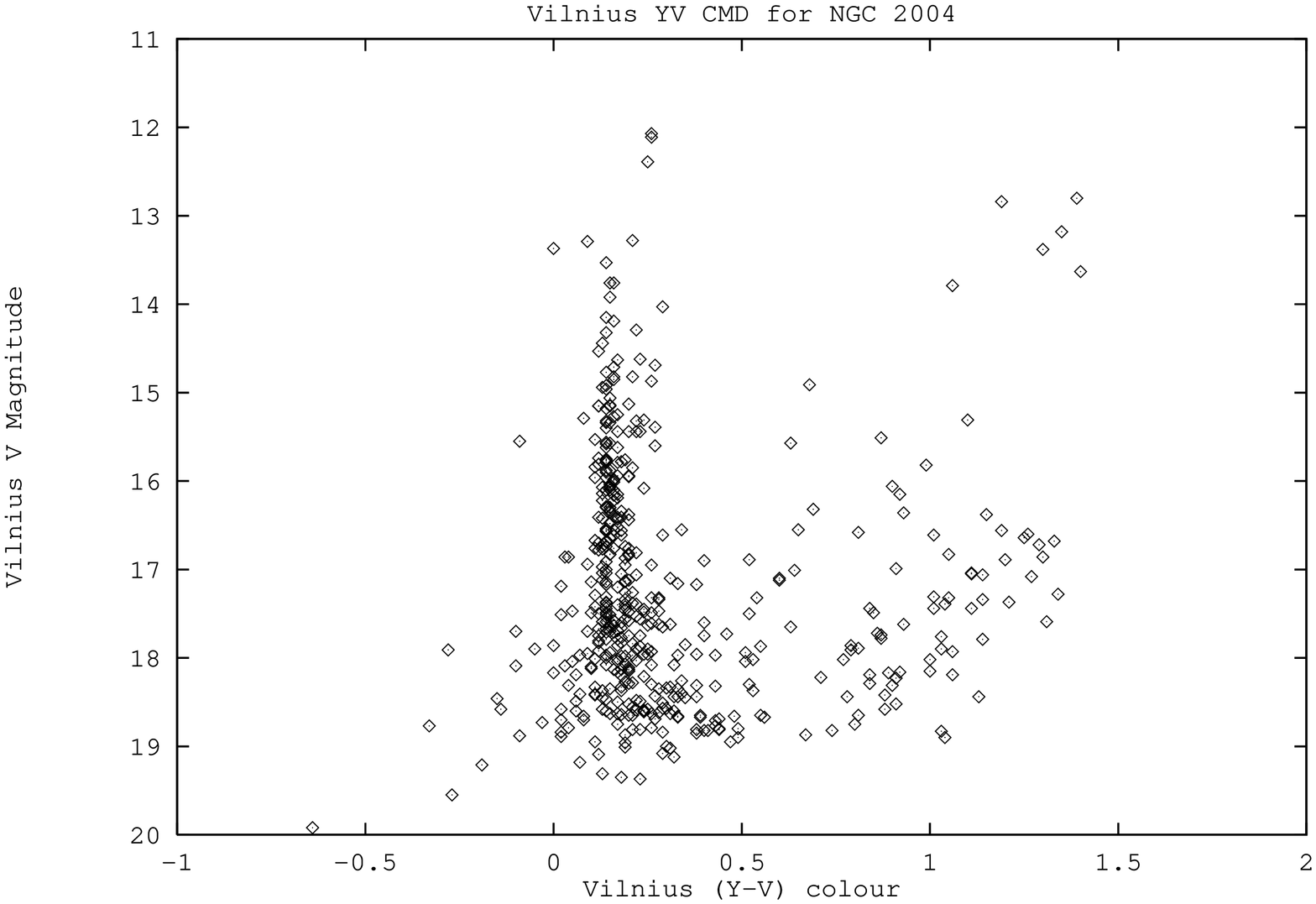,height=7cm,width=7cm,bbllx=20mm,angle=-90}
	}
\subfigure[]
 	{
	\psfig{figure=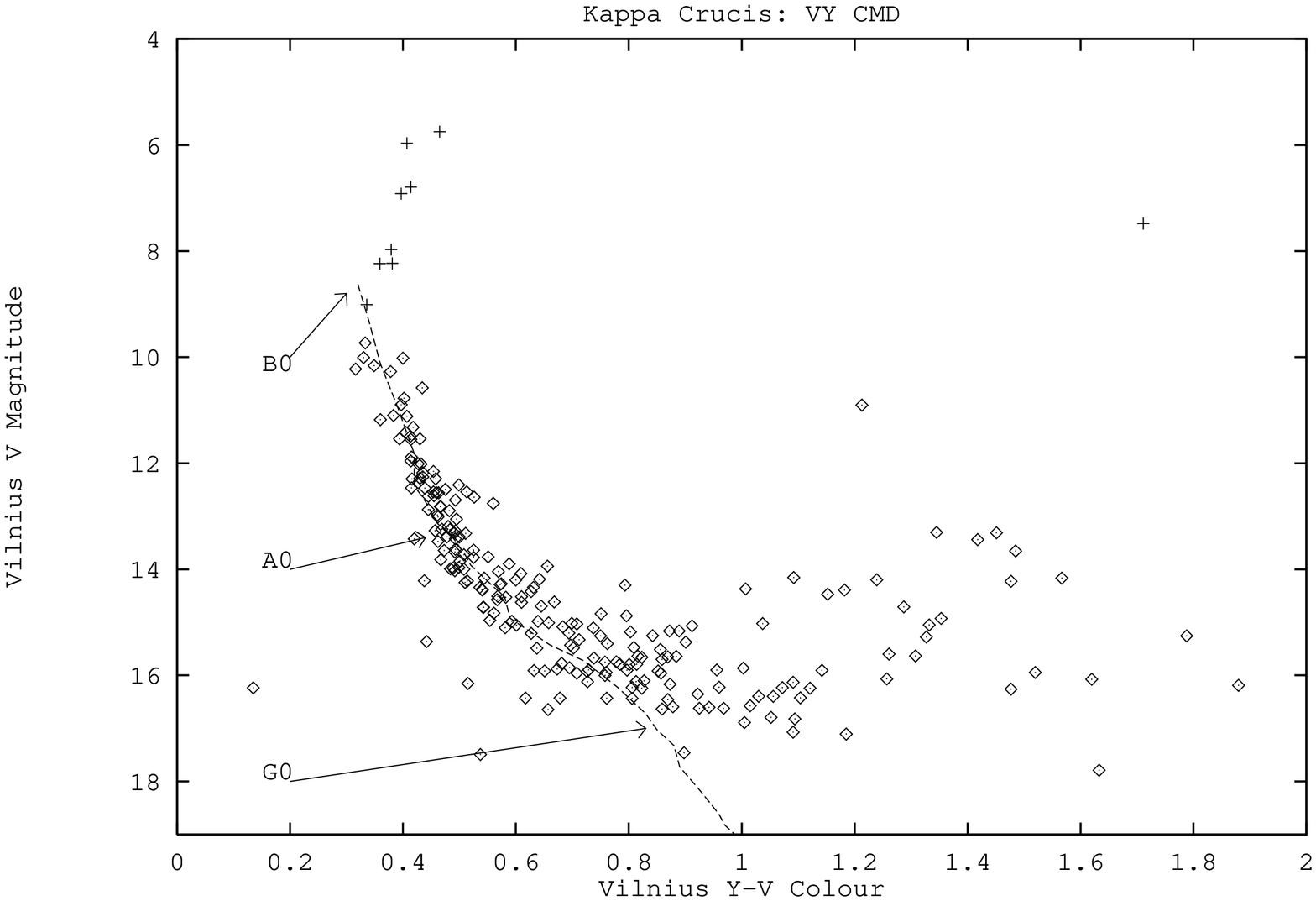,height=7cm,width=7cm,bbllx=20mm,angle=-90}
	}
\end{tabular}

\caption{ {\bf Colour Magnitude Diagrams for NGC~2004 and NGC~4755:}
Subfigure (a) is the standardized, aperture corrected Vilnius YV CMD
for the LMC cluster NGC 2004.  Subfigure (b) is the YV CMD for
NGC~4755 standardized by direct comparison with established Vilnius
secondary standards in the field of the CCD Images. The plus symbols
are photo-electric observations taken from Forbes~(1994). The dashed
line is the Main Sequence as given in Strai\u{z}ys (1992a), taking
into account a 11.85 mag distance modulus and interstellar reddening
E(B-V) of 0.40 mag (Kjeldsen~\& Frandsen 1991). A similar fit will be
made to NGC~2004 once E(B-V) is determined from a colour-colour
diagram. Both CMDs were derived using the IRAF implementation of the
``Classic DAOphot'' (Stetson 1987) crowded field reduction software.
\label{figure:one}}
\end{figure}

NGC 2004 is a young ($ \sim 8 $ x $ 10^7 $ years according to Hodge
1983) populous star cluster in the Large Magellanic Cloud, making it
an attractive object for the study of stellar evolution models.
Several studies have derived Johnson BV Colour Magnitude Diagrams
(CMDs) for it, including Bencivenni~{\em et al.} (1991), and Balona~\&
Jerzykiewicz (1993). Figure~\ref{figure:one}(a) gives the Vilnius YV
CMD based on 19.1 minute exposures standardized using observations of
9 different primary standards over an airmass range of 1.5 to 2.3. The
unusual exposure lengths are due to the Photometrics Acquisition
System running 4.3\% fast for some exposure commands (Tobin 1991).
Unfortunately, clouding in the early morning led to other collected
observations of standards being rejected.  This development of cloud
took place during the NGC~4755 observations.  The standardization
equations: \[ {\rm V_o = V + ( 18.00 \pm 0.05 ) + ( \: ( -0.071 \pm
0.024 ) \: x \: 	 ( Y - V )\: ) + ( \: ( 0.262 \pm 0.028 ) \: x
\: Airmass ) } \] and \[ {\rm Y_o = Y + ( 18.08 \pm 0.05 ) + ( \: (
-0.028 \pm 0.021 ) \: x \: 	( Y - V ) \: ) + ( \: ( 0.273 \pm
0.029 ) \: x \: Airmass ) } \] were obtained using the IRAF Photcal
package, and are only preliminary. They will be recalculated when
values for the secondary standard BS4293 (which was observed at low
airmass several times early in the night) are produced by the
standards programme.  The root mean squares for both the
transformation equations are 0.02 mag. The transformations for the U
and P observations are currently being derived.

The major features of this CMD are the near vertical and well defined
Main Sequence (MS), a few evolved stars in a red giant clump around V
$ \sim $ 13 and $ {\rm ( Y \: - \: V ) \: \sim \: 1.4 } $, and the
field evident at the fainter magnitudes to the red of the MS. These
features, and their relative densities, are identical to those seen in
the literature (see e.g.\ Elson 1991 for a BV CMD acquired with a 1m
telescope equipped like the MJUO 1m). The stars to the blue of the MS
are artifacts of the crowded field reduction and the poor observing
conditions. The widening of the MS towards the fainter magnitudes is a
good reflection of the photometric errors (see Mateo~\& Hodge 1987).
Direct comparison between the CMD presented and literature Johnson BV
CMDs is aided by the fact that the Vilnius V coincides with the
Johnson V magnitude for all unreddened or slightly reddened stars of
spectral types O to K and all luminosity classes (p489, Strai\u{z}ys
1992a).

\subsection{NGC 4755}

It was originally intended to use the secondary standards in this
cluster, in conjunction with the low altitude primary standards, to
derive the transformation equations. When observed, the cluster
standards were at low airmasses. Unfortunately, the above mentioned
formation of cloud prevented this, forcing the use of a direct
transformation based on the ``on chip'' local standards.  The root
mean square errors for the Y and V transformations are large, being
0.03 and 0.05 mag respectively.  The large uncertainties are at least
partially due to the brighter, and more frequently measured, standards
being saturated or approaching pixel saturation, and so the fainter
and less frequently measured stars being used.  The standard
deviations, and their uncertainties, based on the formal errors of the
standard stars themselves (as given by Forbes 1994) are 0.04
and 0.04 mag for the Y and V passbands respectively.
Figure~\ref{figure:one}(b) shows a YV CMD based on 3.2 and 0.8 minute
exposures.  It agrees well with the CMDs of Shobbrook (1984), Dachs~\&
Kaiser (1984), Slettebak (1985), and Frandsen, Dreyer~\& Kjeldsen
(1989). The unusually placed star at V $ \sim $ 11 and Y-V $\sim 1.2 $
is star number 104 in Dachs~\& Kaiser (DK), with a (V, B-V) of (11.03,
1.57), which they excluded as an unlikely member of the cluster. The
bright red star ( V$ \sim 7.5$, Y-V$ \sim 1.7 $) is designated in the
DK scheme as D.  This star ($\kappa $ Crucis itself) was used as the
centre of the Images taken, since it is effectively in the centre of
the cluster.

\section{Discussion}

This preliminary work has demonstrated that the recently acquired 28mm
diameter Vilnius filters allow the use of the MJUO CCD image
acquisition system, and the established data reduction pathway at VUW,
to produce Vilnius photometry in line with the literature.  An
unanticipated problem was encountered with the substantially different
thicknesses of the Vilnius filters, ranging from 9.79 to 3.26 mm, and
the difficulty of ensuring that both the telescope and the offset
guider were in focus for the individual filters. The focus
of the guider is relative to that of the telescope. When the telescope
was refocused to account for the different thicknesses of the filters,
the guider was taken out of focus and had to be adjusted.  Use of the
offset guider is essential if faint limiting magnitudes are to
obtained with the Vilnius system (see Figure~\ref{figure:two} for the
combined response of the filters and the CCD).  The solution to this
problem is to ensure that all the filters are of the same optical
depth, removing the need to change focus between the filters. It is
intended to use Schott FK5 glass to build the P and X filters up to
the depth of the thickest filter (U), and B270 for the remainder. The
spectral profiles and transmittances of the filters should not be
significantly altered by these additions.

Given that the feasibility of the Vilnius filters with the MJUO CCD
system has been demonstrated, it is intended to use the Vilnius system
in further observations of star clusters, and in particular the
populous LMC ones.

\section{Acknowledgments}

The authors are grateful for generous time allocations at Mount John
University Observatory, to the Vilnius Observatory for supplying the
filter set, the New Zealand Lottery Board for financing the purchase
of the set, to Acorn New Zealand for the loan of a R260 computer on
which some of this work was performed, to the Foundation for Research,
Science and Technology for partial funding of this project in
conjunction with the VUW Internal Research Grant Committee, and to the
anonymous referee for helpful comments.  TB acknowledges partial
support during this study by the inaugural R.H.T.  Bates Postgraduate
Scholarship. This paper was a poster presentation at the 1993 PEP-4
Conference.

\section{References}
\footnotesize
{
\begin{itemize}
\item[] Balona, L.A., \& Jerzykiewicz, M., 1993, {\em Monthly
	Notices of the Royal Astronomical Society}, {\bf 260},
	782.
\item[] Banks, T., 1993, {\em Southern Stars}, {\bf 35}, 33.
\item[] Bencivenni, D., Brocato, E., Buonanno, R., \& Castellani,
	V., 1991, {\em Astrophysical Journal}, {\bf 102}, 137.
\item[] Boyle, R.P., Smriglio, F., Nandy, K., \& Strai\u{z}ys, V., 1990,
	{\em  Astronomy \& Astrophysics Supplement Series}, {\bf 84}, 1.
\item[] \u{C}ernies, K., Mei\u{s}tas, E., Strai\u{z}ys, V., \&
	Jasevi\u{c}ius, V.,  1989,
	{\em Bulletin of the  Vilnius Observatory}, {\bf 84}, 9.
\item[] \u{C}ernies, K., \& Jasecvi\u{c}us, V., 1992,
	{\em Baltic Astronomy}, {\bf 1}, 83.
\item[] Dachs, J., \& Kaiser, D., 1984, {\em Astronomy \& Astrophysics
	Supplement Series}, {\bf 58}, 411.
\item[] Dodd, R.J, Forbes, M.C., \& Sullivan, D.J, 1993,
	{\em Stellar Photometry --- Current Techniques and Future Developments},
	IAU Colloquium \#136,
	Eds: C.J. Butler \& I. Elliot, Cambridge University Press, Cambridge, 51
\item[] Elson, R.A.W., 1991, {\em Astrophysical Journal Supplement Series},
	{\bf 76}, 185.
\item[] Forbes, M.C., 1993, {\em Southern Stars}, {\bf 35}, 69.
\item[] Forbes, M.C., 1994, {\em Unpublished PhD Thesis}, Victoria University
of
	Wellington.
\item[] Forbes, M.C., Dodd, R.J., \& Sullivan, D.J., 1993, {\em Baltic
	Astronomy}, {\bf 2}, 246.
\item[] Frandsen, S., Dreyer, P., \& Kjeldsen, H., {\em Astronomy \&
	Astrophysics}, {\bf 215}, 287.
\item[] Hodge, P.W., 1983, {\em Astrophysical Journal},
	{\bf 264}, 470.
\item[] Kjeldsen, H., \& Frandsen, S., 1991, {\em Astronomy~\& Astrophysics
Supplement
	Series}, {\bf 87}, 119.
\item[] Mateo, M., \& Hodge, P.W., 1987, {\em Astrophysical Journal},
	{\bf 320}, 652.
\item[] Moore, C.H., 1974, {\em Astronomy \& Astrophysics Supplement Series},
	{\bf 15}, 497.
\item[] Shobbrook, R.R., 1984, {\em Monthly Notices of the Royal Astronomical
	Society}, {\bf 206}, 273.
\item[] Slettebak, A., 1985, {\em Astrophysical Journal Supplement Series},
	{\bf 59}, 769.
\item[] Stetson, P.B., 1987, {\em Publications of the Astronomical Society
	of the Pacific}, {\bf 99}, 191.
\item[] Strai\u{z}ys, V., 1992a, {\em Multicolour Photometry},
	Volume 15 of ``Astronomy and Astrophysics Series'',
	Pachart Publishing House, Tucson, Arizona.
\item[] Strai\u{z}ys, V., 1992b, {\em Baltic Astronomy},
	{\bf 1}, 107.
\item[] Strai\u{z}ys, V., \& Sviderskiene, Z., 1972,
	{\em Astronomy \& Astrophysics}, {\bf 17}, 312.
\item[] Tobin, W., 1989, in {\em ``Recent Developments of Magellanic
	Clouds Research''}, Eds: K.S. de Boer, F. Spite, and G.
	Stanska, Observatoire de Paris, 177.
\item[] Tobin, W., 1991, {\em Mount John CCD System and Performance
	Note \# 7}.
\item[] Tobin, W., 1992, {\em Southern Stars}, {\bf 34(8)}, 421.
\item[] Wells, D. C., Griesen, E. W., \& Harten, R. H., 1981,
	{\em Astronomy \& Astrophysics Supplement Series}, {\bf 44}, 363.
\item[]  Zdanavi\u{c}us, K., S\={u}d\u{z}ius, J., Sviderskien\.e, Z.,
	 Strai\u{z}ys, V., Burna\u{s}ov, V., Drazdys, R.,
	Bartkevi\u{c}ius, A., Kakaras, G., Kavalianskaite, G., \&
	Jasevi\u{c}ius, V., 1969 {\em Bulletin of the Vilnius Observatory},
	{\bf 26}, 3.
\end{itemize}
}
\end{document}